\documentclass[12pt]{article}

\usepackage[utf8]{inputenc}
\usepackage[T1]{fontenc}
\usepackage{hyperref}
\usepackage{geometry}
\usepackage{setspace}
\usepackage{amsmath,amssymb}
\usepackage[superscript]{cite}

\begin{document}
\thispagestyle{empty}

\begin{center}
{\Large \textbf{Dispersion in Analogue Gravity}}\\

Eren Erberk Erkul\textsuperscript{*} and Ulf Leonhardt\textsuperscript{**}\\[1em]

{\small
\textsuperscript{*}Department of Physics \& Department of Electrical and Electronics Engineering,\\
Middle East Technical University, Ankara 06800, Turkey\\
\texttt{erberk.erkul@metu.edu.tr}\\[1em]

\textsuperscript{**}Department of Physics of Complex Systems,\\
Weizmann Institute of Science, Rehovot 7610001, Israel\\
\texttt{ulf.leonhardt@weizmann.ac.il}\\[1em]
}

\textbf{Dated:} March 30, 2025 \\ 
\end{center}

\begin{center}
\textbf{Abstract}
\end{center}

\begin{center}
\begin{minipage}{0.85\textwidth}
\normalsize
Analogue models of gravity, from Newton to Unruh, have evolved from simple fluid--mechanical models to sophisticated modern experiments. They have shown the robustness of Hawking radiation and highlighted the role of dispersion for quantum fields in curved space. We speculate whether some underlying dispersion may also play a role in explaining the cosmological constant and in resolving the cosmological tensions.
\end{minipage}
\end{center}

\clearpage
\pagenumbering{arabic}

\section*{Act I: Simulating the Curvature}
Analogue gravity, in some form, can be traced back to Isaac Newton. Beyond his famous inverse-square law, Newton speculated about an “aether” or medium that might underlie gravitational phenomena. He initially considered fluid--mechanical models to account for Kepler's laws \cite{Kochiras2021}. In a 1665 letter to Henry Oldenburg and, later, to Robert Boyle \cite{Burtt1924}, Newton wrote:
\begin{quote}
“Gravity is the result of ‘a condensation causing a flow of ether with a corresponding thinning of the ether density associated with the increased velocity of flow.’”
\end{quote}
Although he ultimately formulated gravity as an instantaneous force and famously declined to propose any physical mechanism for it in his \emph{Principia} \cite{Newton1687}, the idea that spacetime could behave like a material medium never entirely disappeared.

In the early 20th century, Wilhelm Gordon reintroduced a more precise version of this concept with the tools of general relativity by showing that light propagation in a moving dielectric can be reinterpreted as propagating in a curved spacetime \cite{Gordon1923}. One starts with the usual Maxwell equations in a medium in Minkowski space. Let the medium have the local isotropic refractive index \(n(\mathbf{r}, t)\) and move with the four--velocity field \(u^\alpha\). One finds that the Maxwell equations in the medium are equivalent to the Maxwell equations in a spacetime with metric tensor
\begin{equation}
g_{\alpha\beta}= \eta_{\alpha\beta} - \left(1-n^{-2}\right) u_\alpha\, u_\beta \,.
\label{eq:gordon}
\end{equation}
Conversely, a spacetime geometry with metric (\ref{eq:gordon}) can be viewed as a moving medium. This is true for arbitrary spacetime geometries, not only for Gordon's metric (\ref{eq:gordon}): they all act on light like effective --- possibly nonisotropic --- moving media \cite{Plebanski1960}, and so does gravity, much like Newton initially thought.\footnote{This discussion applies to bosonic analogues (e.g., phonons/photons), while fermionic analogues require a tetrad structure~\cite{Volovik2023Prism}.}

Several decades later, the next breakthrough came with William Unruh's insight that fluid flows exceeding their local wave speed can mimic black hole horizons \cite{Unruh1981}. He pointed out that when a flow transitions from subsonic to supersonic, phonons are trapped in the supersonic region—exactly as light is trapped inside a black hole. He imagined an experiment with superfluid liquid helium to test Stephen Hawking's theoretical prediction \cite{Hawking1975} that horizons radiate and calculated the rate of Hawking radiation of ``sonic horizons''. 

Unruh's analogy gained prominence with the resolution of the trans-Planckian problem. In the standard Hawking derivation \cite{Hawking1975} modes near the horizon get exponentially blueshifted, implying arbitrarily large frequencies beyond the scale of our current theory. Hawking radiation appears to have come from regimes where we cannot trust physics anymore. Jacobson and Unruh showed \cite{Jacobson1991,Unruh1995,Jacobson1996} that realistic dispersive effects at high frequencies, far from invalidating the effect, actually preserve the Hawking-like radiation. Replacing the linear dispersion relation \(\omega = c\,k\) by a higher-order one like
\begin{equation}
\omega^2 \;=\; c^2\,k^2 \;+\;\alpha\,k^4 \;+\;\dots
\label{eq:dispersion}
\end{equation}
which, though not universal~\cite{RibeiroFischer2023}, prevents unbounded frequency growth.
 Numerical \cite{Unruh1995} and analytic work \cite{Jacobson1991,Jacobson1996} indicated that once a horizon structure is present, the emission of Hawking radiation remains robust. The field of analogue gravity was born, had shown its promise, but was still an obscure subject at the fringe of the mainstream. This changed in the era of experiments. 

In 2000 two papers \cite{Leonhardt2000,Garay2000} showed how to connect Unruh's and Jacobson's analogue with modern experimental tools. None of the original ideas \cite{Leonhardt2000,Garay2000} were immediately applicable; they were still Gedanken experiments --- thought experiments --- in want of more thoughts from both general relativity and experimental physics, but they paved the way to real experiments \cite{Philbin2008,Rousseaux2008,Faccio2010,Weinfurtner2011,Rousseaux2016,Steinhauer2019,Drori2019,Oberthaler2022,Steinhauer2022,Weinfurtner2023}. These experiments worked beyond the wildest dreams of the theorists at the time --- as one of the authors vividly recalls --- in regimes never thought to produce the analogue of Hawking radiation, but they did. For example, in optics \cite{Drori2019} a $10^{-3}$ modification of the effective flow speed, an order of magnitude less than the dispersion of $10^{-2}$, and acting for a mere $10^{-14}\mathrm{s}$ is enough to trigger stimulated Hawking radiation in the ultraviolet. Taken together, the experiments \cite{Philbin2008,Rousseaux2008,Faccio2010,Weinfurtner2011,Rousseaux2016,Steinhauer2019,Drori2019,Oberthaler2022,Steinhauer2022,Weinfurtner2023} all demonstrate that once a horizon is formed --- where the flow speed matches or exceeds wave propagation speed ---Hawking-like emission persists. The details of high-frequency dispersion do not destroy the effect; instead, they are an essential ingredient for the role of the horizon in the generation of Hawking radiation. This collective evidence strongly supports the idea that horizon--induced phenomena are fundamentally robust.

\newpage

\section*{Act II: Defects in the Curvature}
\label{sec2}

We have seen the remarkable resilience of Hawking-like radiation in analogue experiments, even when the experimental regimes differ dramatically from theoretical black holes. Hawking radiation persists even in highly dispersive media. This naturally provokes deeper questions: can these analogues illuminate phenomena beyond classical general relativity? Are these dispersive phenomena hinting at underlying features of gravitation that depart from the traditional continuum picture? One compelling possibility is that dispersion itself is woven into the fabric of spacetime at high energies, departing from the smooth manifold structure.

One case where this might matter is the notorious problem of the cosmological constant \cite{Weinberg1989} $\Lambda$ (also known as ``dark energy'').  The cosmology community agrees that $\Lambda$ ought to be the contribution of the quantum vacuum, because it is the only source term of gravity left in the absence of all other fields and because it is invariant under all coordinate transformations, acting as a universal, invariant background, as the vacuum should. But how exactly does the quantum vacuum produce $\Lambda$? To illustrate what the problem --- and its potential solution --- is, consider the following argument based on a simple dimensional analysis. The vacuum energy must go with $\hbar c$. To turn $\hbar c$ into an energy requires the division by a length and to turn $\hbar c$ into an energy density $\varepsilon$ requires a length to the power of four. For the bare vacuum energy that length is the cutoff of the theory. Jacobson's thermodynamic derivation of Einstein's equations \cite{Jacobson1995} and the Bekenstein entropy of black holes \cite{Bekenstein1973} (with Hawking's prefactor \cite{Hawking1975}) all suggest that $2\ell_\mathrm{P}$ gives the order of that length where $\ell_\mathrm{P}$ is the Planck length:
\begin{equation}
\ell_P \;=\;\sqrt{\frac{\hbar G}{c^3}} \,.
\end{equation}
This would give an energy density $\varepsilon$ that exceeds the actual $\varepsilon_\Lambda$ by a factor in the order of $10^{120}$. On the other hand, if the  vacuum energy is renormalized according to the established quantum field theory in curved space \cite{BD1982} the length scale in $\varepsilon$ is the characteristic scale $\ell$ of cosmic evolution, the Hubble parameter $H$ multiplied by $c$, which would give $10^{-120}$ too little. The correct order of magnitude lies --- logarithmically --- in the middle, as one sees from the Friedmann equation:
 \begin{equation}
H^2 = \frac{4\pi G}{3c^2}\,\varepsilon \quad\implies\quad \varepsilon\propto \frac{\hbar c}{\ell_\mathrm{P}^2\ell^2} \quad\mbox{for}\quad \ell=cH.
\end{equation}
The established theory \cite{BD1982} assumes the validity of the equivalence principle for all scales, which excludes dispersion of the type (\ref{eq:dispersion}) and requires renormalization of up to the fourth order. If dispersion is included, only second--order renormalization is needed \cite{Itay2017}. This, and causality, gives vacuum energies of the correct order of magnitude \cite{Leonhardt2019}. Dispersion near the Planck scale, inspired by analogues of gravity, may thus perhaps explain the cosmological term. 

\newpage

\section*{Act III: Beyond the Curvature}
\label{sec:conclusion-empty}

The preceding sections have illustrated how analogue gravity draws a striking parallel between moving media and curved spacetimes. From Newton’s early speculations \cite{Burtt1924} about an aetherlike medium (refined through Gordon’s work on moving dielectrics \cite{Gordon1923}) to Unruh’s sonic horizons \cite{Unruh1981} with subsequent resolution of the trans--Planckian problem \cite{Jacobson1991,Unruh1995,Jacobson1996} and through experimental demonstrations \cite{Philbin2008,Rousseaux2008,Faccio2010,Weinfurtner2011,Rousseaux2016,Steinhauer2019,Drori2019,Oberthaler2022,Steinhauer2022,Weinfurtner2023}, these developments have sharpened our conceptual intuition of gravity. If we take these dispersive defects of spacetime at face value, they might even be the key to solving some of the long-standing cosmological puzzles. 

Apart from the ``elephant in the room'', the cosmological constant $\Lambda$, the latest major problem of astrophysics is the cosmological tensions \cite{Tensions2022}. Even if $\Lambda$ is accepted as a constant of nature without questioning, there are increasing tensions between the $\Lambda$ Cold Dark Matter model of cosmology and astronomical data. The measured Hubble constant $H_0$ disagrees \cite{Riess2022} with the $H_0$ predicted from the model with parameters fitted to the measured fluctuations of the Cosmic Microwave Background, and so do other observations (to a lesser extent) \cite{Tensions2022}. Although never is anything fully settled in cosmology \cite{April1} the problem is clearly understood and urgently needs a solution. The theory of the quantum vacuum in dispersive spacetime \cite{Leonhardt2019}  suggests a simple solution: it predicts \cite{LandauLeonhardt2024} that quantum fluctuations reduce the weight of matter and radiation as if they were partially buoyant in the ``sea of the vacuum”. Quantities that do depend on their actual masses are not affected, such as the fluctuations of the Cosmic Microwave Background (to a good approximation) but the expansion history of the universe is, This disagreement might very well be the root cause of the cosmological tensions. Due to dispersion near the Planck scale and its influence on the quantum vacuum, the smallest of scales might appear on the grandest of stages,  the Planck scale written on the stars \cite{LandauLeonhardt2024}.

Throughout history, each significant leap in our understanding of gravity (from Newton’s mechanical viewpoint to Einstein’s geometric perspective) has been guided by powerful analogies. It now appears likely that future breakthroughs will also draw upon insights from analogue models, which points to deeper underlying structures beyond the apparent spacetime continuum, raising the question of how far these analogies can be extended and whether fully self--consistent Einstein equations can indeed be encoded in the language of a medium \cite{ErkulLeonhardt2025}
. Physics endeavors to present a unified portrayal of reality, but our understanding is inevitably entangled with the analogies that inspired it. And, in the end, who can definitively say where an analogue system ceases to be a mere analogy and becomes an essential guide to the deeper layers of gravitation?

\section*{Acknowledgements}

We thank Uwe Fischer and Grigory Volovik for helpful comments.\
E.E.E.\ gratefully acknowledges support of the the Weizmann Institute of Science.\
U.L.\ is supported by the Murray B. Koffler Professorial Chair.

\end{document}